\documentclass[11pt, titlepage]{article}
\usepackage[utf8]{inputenc}
\usepackage[english]{babel}
\usepackage{csquotes}
\usepackage[T1]{fontenc}
\usepackage[margin=2cm]{geometry}
\usepackage{appendix}
\usepackage{nameref}

\usepackage{amsmath}
\usepackage{amsfonts} 
\usepackage{mathtools}
\usepackage{amsthm}
\usepackage{physics}
\usepackage{graphicx}
\usepackage{float}
\usepackage[font=scriptsize,labelfont=bf]{caption}
\usepackage{siunitx}
\usepackage{chemformula}
\usepackage{xcolor}
\usepackage{authblk}
\usepackage{textgreek}

\usepackage{comment}

\usepackage{verbatim}

\usepackage{ragged2e}
\usepackage{tabularx}

\usepackage{hyperref}
\hypersetup{
    colorlinks=true,
    linkcolor=blue,
    filecolor=magenta,      
    urlcolor=blue,
    citecolor=blue,
}

\usepackage[backend=bibtex,
            natbib=true,
            sorting=none,
            sortcites=true,
            maxbibnames=6,
            minbibnames=3,
            maxcitenames=2,
            mincitenames=1,
            hyperref=true,
            giveninits=true,
            terseinits=true]{biblatex}
\DeclareNameAlias{sortname}{family-given}
\DeclareNameAlias{default}{family-given}

\DeclareBibliographyDriver{article}{%
  \printnames{author}
  \newunit\newblock
  \printfield{title}
  \newunit\newblock
  \printfield{journaltitle}
  \setunit{\addspace}
  \printfield{year}
  \setunit{\addsemicolon}
  \printfield{volume}
  \iffieldundef{number}{}{\printtext[parens]{\printfield{number}}}%
  \setunit{\addcolon}
  \printfield{pages}
  \newunit
  \printfield{doi}
  \finentry}

\DeclareFieldFormat[article]{title}{#1}
\DeclareFieldFormat[article]{pages}{#1}
\DeclareFieldFormat[article]{journaltitle}{#1\isdot}
\DeclareFieldFormat[article]{doi}{\url{#1}}

\DeclareBibliographyDriver{inbook}{
    \printnames{author}
    \newunit\newblock
    \printfield{title}
    \newunit\newblock
    In:
    \printnames{editor}
    \addcomma\addspace editors.
    \newunit
    \printfield{booktitle}
    \newunit
    \printlist{location}
    \setunit{\addcolon\addspace}
    \printlist{publisher}
    \setunit{\addsemicolon\addspace}
    \printfield{year}
    \newunit
    \printfield{pages}
    \newunit
    \printfield{doi}
    \finentry
}
\DeclareFieldFormat[inbook]{title}{#1}
\DeclareFieldFormat[inbook]{booktitle}{#1}
\DeclareFieldFormat[inbook]{pages}{p. #1}

\DeclareBibliographyDriver{misc}{
    \printnames{author}
    \setunit{\addcomma\addspace}
    \printfield{title}
    \newunit
    \printfield{url}
    \setunit{\addcomma\addspace}
    \printfield{year}
    \setunit{\addspace}
    \printfield{note}
    \finentry
}
\DeclareFieldFormat[misc]{title}{#1}
\DeclareFieldFormat[misc]{url}{\url{#1}}
\DeclareFieldFormat[misc]{note}{(#1)}

\usepackage[ruled,vlined,linesnumbered]{algorithm2e}

\usepackage{booktabs}
\usepackage{rotating}
\usepackage{multirow}
\usepackage[flushleft]{threeparttable}

\usepackage{tabularx}

\providecommand{\keywords}[1]{\textbf{\textit{Keywords---}} #1}

\begin{document}
\begin{center}{\huge\bf Optimal monophasic, asymmetric electric field pulses for selective transcranial magnetic stimulation (TMS) with minimised power and coil heating}

\vspace{12pt}

{
{Ke Ma},
{Andrey Vlasov},
{Zeynep B. Simsek},
{Jinshui Zhang},
{Yiru Li},
{Boshuo Wang},
{David L. K. Murphy},
{Jessica Y. Choi},
{Maya E. Clinton},
{Noreen Bukhari-Parlakturk},
{Angel V. Peterchev},
{Stephan M. Goetz}
}
\end{center}

\vspace{12pt}

\textit{Background}: Transcranial magnetic stimulation (TMS) with asymmetric electric field pulses, such as monophasic, offers directional selectivity for neural activation but requires excessive energy. Previous pulse shape optimisation has been limited to symmetric pulses or heavily constrained variations of conventional waveforms without achieving general optimality in energy efficiency or neural selectivity. \\
\textit{Objective}: We sought to develop a minimally constrained optimisation framework for identifying energy-efficient asymmetric TMS pulses with directional selectivity of neural stimulation. \\
\textit{Methods}: {We implemented a novel optimisation framework that incorporates neuron model activation constraints and flexible control of pulse asymmetry. The optimised waveforms were experimentally validated against conventional and previously optimised pulses. We measured motor thresholds for conventional pulses as well as one of the optimised unidirectional rectangular (OUR) pulses and compared its MEP latency for anterior--posterior (AP) and posterior--anterior (PA) electric field directions in six healthy human subjects.}  \\
\textit{Results}: The optimised electric field waveforms had leading phases with a time constant of {(280\,$\pm$\,15)\,{\textmu}s} (mean\,$\pm$\,SD) and near-rectangular main stimulation phases. They achieved up to 92\,\% and 88\,\% reduction in energy loss and thus coil heating respectively compared to conventional monophasic pulses and previously improved monophasic-equivalent pulses. In the human experiments, OUR pulses showed similar motor thresholds to monophasic pulses in both AP and PA directions whilst achieving significantly lower energy loss, particularly in the AP direction. Moreover, there was a significant MEP latency difference of (1.79\,$\pm$\,0.41)\,ms (mean\,$\pm$\,SE) between AP and PA direction with OUR pulses, which suggests directional selectivity. \\
\textit{Conclusion}: Our framework successfully identified highly energy-efficient asymmetric pulses for directionally-selective neural engagement. These pulses can enable selective rapid-rate repetitive TMS protocols with reduced power consumption and coil heating, with potential benefits for precision and potency of neuromodulation.  \\

\section{Introduction}
Transcranial magnetic stimulation (TMS) is a widely used noninvasive brain stimulation technique based on electromagnetic induction \cite{barker1985non, polson1982_pulse_energy}. TMS pulses are typically classified into monophasic (with an asymmetric and more unidirectional electric field) and biphasic (with a more symmetric and thus bidirectional electric field), which are widely available in commercial devices \cite{rossini2015_IFCN,kammer2001_waveforms}. The currents through the stimulation coil, which directly touches the patient, reach thousands of amperes, require substantial power, and cause excessive heating of the coil \cite{wagner2007_TMS_review,goetz2017_TMS_Review}. The high power requirements and safety restrictions on coil heating complicate device technology, constrain device flexibility, and necessitate active coil cooling systems to enable repetitive TMS (rTMS) protocols \cite{nielsen1995_oil_cooled_coil}.

At present, biphasic pulses can be generated more efficiently with lower coil heating and are the dominant shape in high-frequency rTMS, whereas monophasic pulses serve primarily in single-pulse and low-frequency stimulation \cite{goetz2017_TMS_Review,rossini2015_IFCN,polson1982_pulse_energy}. However, monophasic and generally asymmetric electric field pulses are considered to be more selective in recruiting cortical neurons than biphasic pulses \cite{niehaus2000_pulse_efficacy, kammer2001_waveforms, sommer2006_waveform_motor_cortex, dostilio2016effect}. For instance, monophasic pulses show strong directional selectivity in their effect on cortical neurons \cite{di1998_monophasic_volleys, di2001_monophasic_volleys}, whereas this directional effect is much less pronounced with biphasic pulses \cite{di2001_biphasic_volleys, weise2023directional}. Measurements of motor response latencies also indicate high selectivity of asymmetric electric field pulses \cite{dostilio2016effect, sommer2006_waveform_motor_cortex, goetz2016_Enhancement_RecMono}. Moreover, numerous studies suggested that asymmetric electric field pulses offer higher neuromodulatory efficacy, e.g., inducing stronger plasticity, compared to biphasic pulses with a symmetric electric field \cite{sommer2002_mono_effective, arai2007_mono_effective, taylor2007_MONO, goetz2016_Enhancement_RecMono, nakamura2016_variability_QPS, wendt2023_iTBS_monoVSbi}.

Due to their superior selectivity and neuromodulation efficacy, monophasic pulses appear attractive for clinical therapy. However, conventional monophasic devices are inefficient and ill-suited for rTMS because their high pulse energy loss not only exacerbates the required supply power and coil heating but also limits the maximum duration of the stimulation session and the pulse repetition rate \cite{weyh2005_TMS_heating_equation,rossi2009safety}. We developed a modular pulse synthesiser TMS (MPS-TMS) device that can flexibly and efficiently synthesise arbitrary pulse shapes, which, for the first time, enables the delivery of pulses designed for specific objectives such as reduced coil heating and/or increased neural activation selectivity \cite{li2022modular,goetz2012circuit,zhang2025asymmetric,zhang2025high}.

{
Prior work on TMS pulse optimisation has several significant limitations. An unconstrained optimisation framework generated near-rectangular electric field waveforms with improved efficiency and reduced coil heating, but the pulses had mostly symmetric amplitudes in both directions and were therefore not selective \cite{goetz2013_biphasic_opt,zhang2023multi}. Another approach constrained the root-mean-square error (RMSE) of the optimised electric field waveforms compared to that of conventional monophasic pulses, which determined how much the optimised electric field pulses could diverge from the conventional monophasic template \cite{wang2022optimized}. The optimisation starts from the conventional monophasic current waveform as an initial state and improves energy efficiency by perturbing the current waveform whilst retaining the electric field waveform similarity to the original monophasic electric field pulse shape within certain limits \cite{wang2022optimized}. Although this approach enhances pulse efficiency, it has three main limitations. First, the selection of the template depended on existing waveforms; it potentially overlooks a wide range of pulse shapes that could offer even greater asymmetry and still lower energy loss. Second, the selection of the constraints for the allowed deviation of the electric field was arbitrary, and the optimised pulses demonstrated that a larger deviation significantly reduced the asymmetry of the pulse shape \cite{wang2022optimized}. Third, the method generated a pulse better than the reference, but provided no information on how good the pulse is on an absolute scale, i.e., whether the pulse for the given asymmetry level is nearly optimal or still has potential for large further improvement.

To explore and establish the potential of the entire waveform space, this paper presents a minimally constrained optimisation framework for identifying energy-efficient asymmetric TMS pulses that maintain neural stimulation selectivity. In contrast to previous approaches for symmetric pulses or constrained variations of conventional waveforms, this framework explores a wide range of electric-field asymmetry levels to identify globally optimal pulse shapes. We establish the theoretical foundation for the optimisation process, experimentally validate the optimised waveforms against conventional pulses with respect to energy loss, and compare the motor threshold and motor-evoked potential (MEP) latency between two pulse directions as an indicator of neural selectivity. This work advances TMS technology by enabling selective stimulation protocols with reduced power consumption and coil heating, which could potentially increase the efficacy of TMS in clinical and research settings.
}

\section{Optimisation framework}
\subsection{Methods and constraints}
{The optimisation problem aims to minimise energy losses of the TMS current waveforms.} The energy losses of the device and the coil are dominated by Joule heating of the internal resistance of the coil and the device in the first approximation. For a coil current $i(t)$, the energy loss of a pulse is
\begin{equation}
    \int_{\mathbb{R}^+}\!\!\!{}R_\textrm{coil} \cdot i^2(t) \,\dd{t} \, =: \, \mathcal{W},
    \label{equ: energy loss}
\end{equation}
where $R_\textrm{coil} > 0$ is the equivalent intrinsic resistance. $\mathcal{W}$ serves as an indicator of energy losses and coil heating in the following.

\subsubsection{Neuron activation}
{
For the first constraint, the induced electric field must be strong enough to elicit a neural response. The neuron model selection deserves consideration. \citet{goetz2013_biphasic_opt} investigated the sensitivity of pulse optimisation to different neuron models and found that the key features of optimised waveforms remain consistent across various computational neuron models. \citet{wang2022optimized} demonstrated that single- and multi-compartment models have similar relative activation thresholds for different pulse shapes in numerical simulation. In addition, our study concentrates on the excitation dynamics rather than action potential propagation. Thus, we used a previously published nonlinear single-compartment model of a human axon as the target neuron model \cite{goetz2013_biphasic_opt}. Without loss of generality of neuron dynamics, this nominal model incorporates fast sodium, persistent sodium, and dominant slow potassium channels as well as passive leakage; it therefore encompasses the common excitation dynamics of various types of targets, such as motor, corticospinal, or interneuron axons \cite{goetz2013_biphasic_opt, richardson2000modelling, mcintyre2002modeling}. Table S1 in the Supplementary Materials summarises all formulations and parameters. The model was stimulated with an intracellular current injection proportional to the electric field waveform. The coefficient of proportionality represented the coupling strength between the electric field and the neuron model (i.e., activating function) as well as the cellular geometry (axon size). The electric field coupling was set to 10 ({\textmu}A/$\mathrm{cm^2}$)/(V/m) \cite{mainen1996influence}.
}

{
This optimisation framework ensures that the axon does not spike spontaneously and is in the resting steady state before stimulation. The neuron activation constraint is implemented as a binary requirement based on a threshold value for the trans-membrane potential. The neuron model is classified as activated if its normally negative trans-membrane potential exceeds $+10\,\textrm{mV}$; otherwise it is not activated. We selected this positive threshold value to avoid false activation determination. It introduces a substantial safety margin that improves determination tolerance to numerical estimation errors and ensures reliable supra-threshold responses of neuron models. This approach prevents the optimisation from finding pulses that barely reach the activation threshold for the sake of small energy losses.
}

\subsubsection{Voltage limits}
This study identifies the coil current as a real-valued function $i(t)$ over time with zero values at both ends. {Under quasi-static approximations, in which tissue properties are considered linear, resistive, and frequency-independent \cite{wang2024quasistatic}, the induced electric field $\boldsymbol{E}(\boldsymbol{r}, t)$ in the tissue is proportional to the time derivative of the coil current
\begin{equation}
    \boldsymbol{E}(\boldsymbol{r}, t) =  \boldsymbol{k}_E(\boldsymbol{r})\cdot \dv{i(t)}{t}.
\end{equation}
Here, $\boldsymbol{k}_E(\boldsymbol{r})$ is the electric field distribution map with spatial coordinate $\boldsymbol{r}$ for a unit rate-of-change in coil current. In the motor cortex, its magnitude is approximately 1 (V/m)/(A/{\textmu}s) \cite{aberra2020simulation}. The coil voltage is related to the current by $V(t) = L_{\mathrm{coil}}\cdot \dv*{i(t)}{t} + R_{\mathrm{coil}}\cdot i(t)$. For TMS pulses, the rate of change of the coil current, i.e., inductive voltage, primarily dominates the peak coil voltage rather than the resistive voltage. In addition, for typical coil inductance and resistance values, the resistive voltage drop at maximum or minimum coil current (when the rate of change is zero) is smaller. Therefore, the optimisation only considered the coil voltage to be proportional to the time derivative of the coil, i.e., $V(t) = L_{\mathrm{coil}} \cdot \dv*{i(t)}{t}$.} We impose the asymmetry on the electric field and ergo the coil voltage through limiting the maximum positive and minimum negative voltages ($V_\textrm{max}$ and $V_\textrm{min}$) as the second constraint. These values further form a voltage and electric-field asymmetry ratio $r_\textrm{V} = \abs{\frac{V_\textrm{max}}{V_\textrm{min}}} = \abs{\frac{E_\textrm{max}}{E_\textrm{min}}} = r_\textrm{E}$.

{
Without voltage limits, the pulse optimisation will shorten the pulse at the cost of higher peak voltages as well as electric fields, which are known to reduce the energy loss \cite{goetz2012optimization,barker1991_Pulse_Duration_Amplitude,weyh2005_TMS_heating_equation,peterchev2013pulse,Zeng2022}. However, there are practical hardware limits on the maximum coil voltage, including insulation, semiconductor switching device rating, and voltage-dependent switching losses \cite{peterchev2011repetitive,Zeng2022,li2022modular}. We translated the constraint of the voltage limit into a penalty ($U_\textrm{p}$ in Table \ref{tab: optimisation problem}) rather than fragmenting the solution space. The approximately convex penalty allows for a potentially faster return to the valid solution space by presenting a steep but well-defined gradient back once the optimisation exceeds the voltage limits. A multiplier $\lambda$ was assigned to the penalty term, which also includes the units, to ensure compliance with the voltage constraints.
}


\begin{table}[ht]
    \centering
    \caption{Summary of the optimisation problem}
    \resizebox{0.8\linewidth}{!}{
    \begin{threeparttable}
    \begin{tabular}{llll}
        \hline
        \textbf{minimise} & & & $\mathcal{J}(i(t)) = {R_{\mathrm{coil}} \cdot}\int i^2(t) ~ \dd{t} +                                 {\lambda\cdot } \left(\int  U_\textrm{p}(i(t)) ~ \dd{t} \right)^2$ \\
         & & & {where $U_\textrm{p}(i(t)) =
                      \begin{cases}
                        L_{\mathrm{coil}} \cdot\dv{i(t)}{t} - V_\textrm{max}, & \text{if } L_{\mathrm{coil}} \cdot\dv{i(t)}{t} > V_\textrm{max}, \\
                        V_\textrm{min} - L_{\mathrm{coil}} \cdot\dv{i(t)}{t}, & \text{if } L_{\mathrm{coil}} \cdot\dv{i(t)}{t} <  V_\textrm{min}, \\
                        0, & \text{otherwise},
                      \end{cases}$}\\
                      &&& {$R_{\mathrm{coil}}= 10 \ \unit{\mohm}$, $L_{\mathrm{coil}}=10${\,\textmu}H, and $\lambda = 1$\,S/s,}\\ 
        \textbf{over} & & & $i(t) \approx \vb{I} = [I_k]$, $k = 1,2,\cdots,N$, \\
                      & & & within a time window $[0,3]\,\unit{\milli\second}$, \\
        \textbf{subject to} & & & 1. $I_1 = I_N = 0$, \\
                            & & & 2. $\vb{E}_\textrm{intp}$ can evoke an action potential in the neuron model, \\ 
        \textbf{with} & & & $\vb{I}_\textrm{intp} = [I_{\textrm{intp}, j}]$, $j = 1,2,\cdots,M$, \\
                      & & & $\vb{E}_\textrm{intp} = [E_{\textrm{intp}, j}] = |\boldsymbol{k}_E|\cdot [\frac{I_{\textrm{intp}, j} - I_{\textrm{intp}, j-1}}{\Delta{T}}]$, \\  & & & $\Delta{T} = 1\,${\textmu}s, $|\boldsymbol{k}_E|=1\ $(V/m)/(A/{\textmu}s).  \\ \hline
    \end{tabular}
    \begin{tablenotes}[para]
        \textbf{Note:} (a) $U_\textrm{p}$ is the penalty function for the voltage when it exceeds the lower or higher voltage limits. (b) $\vb{I}$ is the current array vector with $N$ elements. (c) $\vb{I}_\textrm{intp}$ is the interpolated current array with a time step $\Delta{t} = 1\,${\textmu}s and a number of sampling points $M$ for $\vb{I}$. (d) $\vb{E}_\textrm{intp}$ is the time derivative of $\vb{I}_\textrm{intp}$.
    \end{tablenotes}
    \end{threeparttable}
    }
    \label{tab: optimisation problem}
\end{table}

\subsection{Implementation}
In this optimisation problem, the current function, $i(t)$, is discretised into a numeric array $\vb{I}$ with $N$ degrees of freedom for use in the numerical optimisation procedure. $\vb{I}$ consists of the parameters of spline curves, i.e., jointed differentiable low-order functions, which are numerically evaluated to $\vb{I}_\textrm{intp}$ with a time step of $\Delta{t} = 1\,${\textmu}s over the entire time window of $3\,\unit{\milli\second}$ with $M=3001$ sampling points for subsequent calculation of the induced electric field waveform $\vb{E}_\textrm{intp}$ (listed in Table \ref{tab: optimisation problem}). This parameterisation translates a set of parameters with finite but variable degrees of freedom $N$ into a highly flexible continuous and differentiable current profile and, after differentiation, into the voltage and the electric field waveforms. The parameterisation is stable and general so that any waveform can be approximated by a sufficiently high value of $N$. The degrees of freedom $N$ can be varied dynamically to refine solutions and allow for higher-frequency features. An existing solution can further be converted into one with different degrees of freedom, e.g., to refine it with higher degrees of freedom. The initial degree of freedom $N$ of each optimisation run is chosen randomly from a range of $[25, 100]$.

We calculated the time derivative of $\vb{I}_\textrm{intp}$ to form the induced electric field waveform $\vb{E}_\textrm{intp}$, which is passed to the single-compartment neuron model as a stimulation input.

Table \ref{tab: optimisation problem} summarises the optimisation problem and its objective function with the coil inductance and resistance set to $10$\,\textmu{H} and $10\,\unit{\mohm}$. $U_\textrm{p}$ is the penalty function for the voltage when it exceeds the lower or higher voltage limits (in units of volts). Due to spatial-temporal separability under quasi-static approximation \cite{wang2024quasistatic}, this optimisation only focuses on the design of the temporal properties of TMS pulses that have a high asymmetry ratio and can stimulate neurons rather than the spatial properties, such as the distribution of the electric field induced by stimulation coils. Therefore, this optimisation procedure does not require the consideration of any specific configurations of stimulation coils and devices as restricted constraints and naturally maintains its generalisation. Nevertheless, instead of arbitrary units, we selected typical values for parameters to establish physically meaningful units. By doing so, the optimised current waveforms are ready to be implemented in experiments after scaling their amplitudes for various TMS equipment.

\subsection{Optimisation procedure}
The optimisation uses a hybrid global--local framework. The global part, a particle swarm here, needs to manage the numerous local minima due to the high nonlinearity of the problem. The local optimisation algorithm uses an interior-point method to speed up convergence to dominant local minima \cite{byrd2000trust}. The local optimiser instances are nested inside the global particle swarm optimiser, where each local optimiser represents a particle.

{
We implemented the particle swarm method to combine the information about the local minima found by the local optimiser instances and update the waveform parameters for them \cite{eberhartnew}. We used inertia parameter $\omega = 0.9$ as well as the gravity parameters $c_1 = 1.2$ and $c_2 = 0.12$ for the particle swarm method. The global framework adaptively controls the number of degrees of freedom. For the first run, the algorithm generates the initial array with an initial degree of freedom $N$ by extracting values from a normal distribution with zero mean and $N$ standard deviation. It increases the degrees of freedom $N$ by $5$ per iteration if convergence is achieved in the current iteration and the objective outperforms the previous one; otherwise, the number of degrees of freedom remains constant. A new global minimum is updated if a local minimum has a lower cost than the current global minimum. In order to expand the optimisation exploration space, after completion of each iteration we introduced random noise, which follows the particle update rule, to the best local minimum as the initial state for the next iteration. A workstation computer (Intel Core i7-12700 CPU) performed each optimisation condition (i.e., different asymmetry ratios) independently. The method was set up in MATLAB (The MathWorks, USA).
}

We tested 18 different combinations of voltage limits and asymmetry ratios (Table S2 in Supplementary Materials), ranging from $r_\textrm{V} = 0.5 = 1\!\!{}:\!\!{}2$ to $r_\textrm{V} = 20\!\!{}:\!\!{}1$, with the highest limit being $V_\textrm{max} = 4000\,\textrm{V}$ and the lowest $V_\textrm{min} = -2000\,\textrm{V}$. The number of degrees of freedom reached up to one thousand during the optimisation process. {In addition, we perform ten independent runs for each combination of voltage limits.} The induced electric field waveforms with $r_\textrm{V} \approx 1\!\!{}:\!\!{}1$ are called symmetric pulses in the following.

\section{Experimental methods}
This study used an MPS-TMS prototype with six modules to synthesise both optimised pulses as well as conventional sinusoidal monophasic and biphasic pulses \cite{li2022modular}. To validate the computational optimisation results and \textit{in vivo} performance of the optimised pulses, we conducted a set of experiments to compare the synthesised and computational waveforms, their neural activation thresholds and energy loss, as well as the directional neural recruitment selectivity of the optimised pulses.

\subsection{Waveform comparisons}
We combined the MPS-TMS device with a standard Cool-B65 coil (MagVenture, Denmark) to generate the optimised asymmetric pulses and the previously improved five monophasic-equivalent pulses with RMSE of $0.25\,\%$, $0.5\,\%$, $1\,\%$, $2\,\%$, and $4\,\%$ with pulse durations of $6\,\textrm{ms}$ \cite{wang2022optimized} for comparison of energy loss. The coil had an inductance $L_\textrm{coil} =11.9\,${\textmu}H and resistance $R_\textrm{coil} = 11.1\,\unit{\mohm}$ at $1\,\unit{\kHz}$. Before each pulse, the internal capacitors of the modules were charged up to $50\,\unit{\volt}$, which could generate an output voltage ranging from $-300\,\unit{\volt}$ to $300\,\unit{\volt}$. The absolute peak voltage of every generated voltage pulse equals the maximum voltage of MPS-TMS output for a peak-matched comparison. We followed the same experimental procedure for measuring optimised pulses as previously suggested in the literature \cite{wang2022optimized}.

To match the amplitude of waveforms between the two studies, we first scaled the electric field waveform of the original monophasic and optimised monophasic-equivalent pulses such that they could successfully elicit an action potential in the neuron model in our study. When generating the asymmetric waveforms experimentally, we also scaled their amplitude so that the amplitude ratio between them and the monophasic-equivalent pulses stayed constant between the modelling and experiments for a threshold-matched comparison.

We calculated the energy losses of the experimentally-recorded waveforms through
\begin{equation}
    \hat{\mathcal{W}} = R_\textrm{coil} \int i^2(t)_\textrm{coil} ~\dd{t},
    \label{equ: calculate real losses}
\end{equation}
where $I_\textrm{coil}$ was the coil current and $R_\textrm{coil}$ represents the coil resistance. The coil resistance was measured with an impedance meter (889A, B\&K Precision, USA). We quantified the change in the loss of the optimised pulses, i.e., loss reduction, as a percentage of a selected reference pulse ($\hat{\mathcal{W}}_\textrm{Ref}$) per
\begin{equation}
    \eta = \frac{\hat{\mathcal{W}} - \hat{\mathcal{W}}_\textrm{Ref}}{\hat{\mathcal{W}}_\textrm{Ref}} \times 100\,\%.
    \label{equ: loss gain}
\end{equation}

{
The commercial TMS coils used in this study were constructed using litz wire \cite{drakaki2022database}, which is commonly employed in high-performance coils to minimise skin and proximity effects. The litz wire minimises frequency-dependent coil resistance effects. Furthermore, the conventional and optimised pulses exhibited similar frequency spectra, with the majority of the energy content concentrated in the low kilohertz range (see Figure S1 in Supplementary Materials for the current spectra). Therefore, the coil resistance is regarded as approximately constant within the pulse frequency range and matched across the different pulse waveforms employed in this study.
}

\subsection{Waveform recordings}
The waveforms, e.g., current, voltage and electric field, were recorded with a $2.5\,\text{GHz}$ sampling rate oscilloscope (MDO3054, Tektronix, USA), followed by a first-order Butterworth low-pass filter with a bandwidth of $200\,\unit{\kilo\hertz}$ to remove measurement noise.

\subsection{Human experiments}
\subsubsection{Participants}
The number of subjects was determined through a power analysis under Cohen's conventional framework (see Section 3 in the Supplementary Materials). Six healthy right-handed volunteers (2 female, 4 male, age = (32.8\,$\pm$\,12.0) years (mean $\pm$ SD), 20--51 years (range)) participated in the study (see Section 4 in the Supplementary Materials for individual details). The participants were screened for contraindications to TMS and neurological and psychiatric disorders. The human study was conducted in accordance with the protocol approved by the Institutional Review Board of Duke University Health System (Pro00112338). All participants completed an informed consent process. Each study session comprised various TMS measurements, including the motor thresholding and MEP latency determination reported here, and participants received $\$50$ per session for their time.

\subsubsection{Experimental setup}
A standard C-B70 coil (MagVenture, Denmark) served for the human experiments. This coil has an inductance $L_\textrm{coil} = 12.1\,${\textmu}H and resistance $R_\textrm{coil} = 13.0\,\unit{\mohm}$ at $1\,\unit{\kHz}$. From the 18 optimised asymmetric pulses, we selected \textrm{V:+2000/--100} with a high asymmetry ratio of $r_\textrm{V} = 20$---named optimised unidirectional rectangular (OUR) pulse in the experimental study---because it has a positive phase which is comparable to that of the conventional monophasic pulse but avoids its substantial negative phase. The MPS-TMS device generated OUR as well as conventional monophasic and biphasic sinusoidal pulses and administered them via the coil with the induced electric field in both posterior--anterior (PA) and anterior--posterior (AP) directions (Figure \ref{fig: human waveforms}(\textbf{A})). Before each pulse delivery, the device charged the internal module capacitors up to $466\,\unit{\volt}$, which could generate an output voltage ranging from $-2.8\,\unit{k\volt}$ to $2.8\,\unit{k\volt}$. The direction refers to the phase of the induced electric field that dominates the stimulation effects for each pulse type: the first phase for monophasic pulses and the second phase for biphasic and OUR pulses \cite{corthout2001transcranial,goetz2016_Enhancement_RecMono}. The pulse direction was switched electronically, without moving the TMS coil. During the experiment, participants were seated in a comfortable chair (Neuronetics, USA) and wore foam earplugs for hearing protection \cite{goetz2015impulse}. Subjects were provided with a pillow to rest their arm on and instructed to relax their limbs.

\subsubsection{Electromyography}
{Electromyography (EMG) was recorded from the first dorsal interosseous (FDI) of the right hand as the target muscle. Since novel TMS pulse shapes were involved, we also recorded EMG from two additional hand muscles (abductor digiti minimi and abductor pollicis brevis) and two forearm muscles (extensor carpi radialis and flexor carpi radialis) on the right and monitored all muscles for safety including afterdischarges and spread of excitation as potential seizure precursors \cite{rossi2021safety, wassermann1998risk}.  Kendall 135 foam electrodes (Cardinal Health, Inc., USA) were placed in a lengthwise bipolar belly--tendon montage.} A BrainAmp ExG amplifier and BrainVision Recorder software (Brain Products GmbH, Germany)  amplified, band-pass filtered (10--1,000 $\unit{\hertz}$), and digitised the EMG signal with a sampling rate of $5\,\unit{\kilo\hertz}$ and $0.5\,${\textmu}V resolution. MATLAB served to process the acquired data. MEPs with suprathreshold activation within 105\,ms before the TMS pulse were marked as facilitated by pre-activation, and the corresponding TMS pulse was redelivered after 6--10\,s. Facilitated MEPs were excluded from all analyses.

\subsubsection{TMS neuronavigation}
A Brainsight neuronavigation system with retroreflective markers (Rogue Research Inc., Canada) served to localise and maintain the TMS coil position relative to the subject's head throughout the study. The subject tracker was attached to the forehead, and the head was registered to the MNI-152 template in Brainsight.

\subsubsection{TMS targeting}
The TMS motor hotspot, corresponding to the largest motor MEPs of the right FDI muscle, was identified using supra-threshold PA OUR pulses delivered at 3--6\,s interstimulus intervals. The muscle twitches were monitored visually and with online EMG. The TMS coil was placed over the hand knob region of the left primary motor cortex, tangentially to the skull and at $45^{\circ}$\! to the mid-sagittal axis. The coil was systematically moved over the hand knob area to identify the location with the largest MEP responses \cite{rossi2021safety,rossini2015_IFCN}. Subsequently, the coil position was maintained within $2.5\,\unit{\milli\meter}$ of the motor hotspot with the help of the neuronavigation system. The TMS coil placement was stabilised with a mechanical coil holder with continuous manual support and adjustment.

\subsubsection{Motor threshold titration}
Resting motor threshold (RMT) is defined as the stimulation strength needed to elicit an MEP of median $50\,${\textmu}V peak-to-peak amplitude in the target muscle at rest. RMT was titrated for three types of pulses (OUR, monophasic, and biphasic) in two directions (posterior--anterior, PA, and anterior--posterior, AP) with an adaptive stochastic approximation method \cite{wang2023three,tms_samt_2024,wang2024samt}. The interstimulus interval was 6--10 s. RMT titrations which resulted in less than two or more than eight suprathreshold ($\geq 50\,${\textmu}V) MEPs over the last ten trials were considered inaccurate and rerun \cite{tms_samt_2024}. RMT determination for the six pulse conditions was conducted simultaneously by interleaving the pulse configurations pseudorandomly to minimise the impact of fluctuations of brain excitability over time.

\subsubsection{MEP latency measurement}
MEP latency was measured for the OUR pulse by applying eight stimuli in PA and AP direction at $110\,\%$ of the respective RMT in an interleaved manner. The inter-stimulus interval was 6--10 s. MEP latency was measured relative to the main phase of the pulses and calculated by applying Bigoni's method \cite{bigoni2022_autolatency} to the filtered FDI EMG data, with results manually verified and corrected where necessary. MEP latency measurement could not be completed for one subject due to their inability to maintain muscle relaxation during the procedure.

\subsection{Statistical analysis}
We employed three mixed-effect models to respectively analyse RMT, energy loss, and MEP latency for different pulse shapes and directions. For RMT and energy loss analyses, we selected \textit{pulse shape} (three levels: monophasic, biphasic, and OUR), \textit{pulse direction} (two levels: PA and AP), and their interaction as fixed-effect variables. The MEP latency analysis was restricted to OUR pulses only; therefore, \textit{pulse direction} served as the sole fixed effect in this model. To account for inter-subject variability, we incorporated the subject as a random intercept term across all three mixed-effects models. We used the \textit{lme4} package (version 1.1-33) in R (version 4.3.1) to estimate the parameters of these mixed-effect models. To test the significance of the fixed-effect variables, we conducted a type-III ANOVA with Satterwaite's method. To compare across the levels of significant factors, we employed estimated marginal means in \textit{emmeans} package (version 1.8.7) with adjustment for multiple comparisons using the Tukey method for post-hoc analysis \cite{searle1980_emmeans,emmeans}. Moreover, we conducted a leave-one-out resampling cross-validation analysis for all mixed-effect models to assess the robustness of our findings. This approach systematically evaluated the model predictive performance by iteratively fitting models using data by excluding one subject and resampling the data through all permutations. The Supplementary Materials (Section 5) provide details of the cross-validation procedure and results. The significance level was set to $\alpha = 0.05$ for all analyses.

\section{Results}

\subsection{Common features of the optimised waveforms}

\begin{figure}
    \centering
    \includegraphics[width=0.5\linewidth]{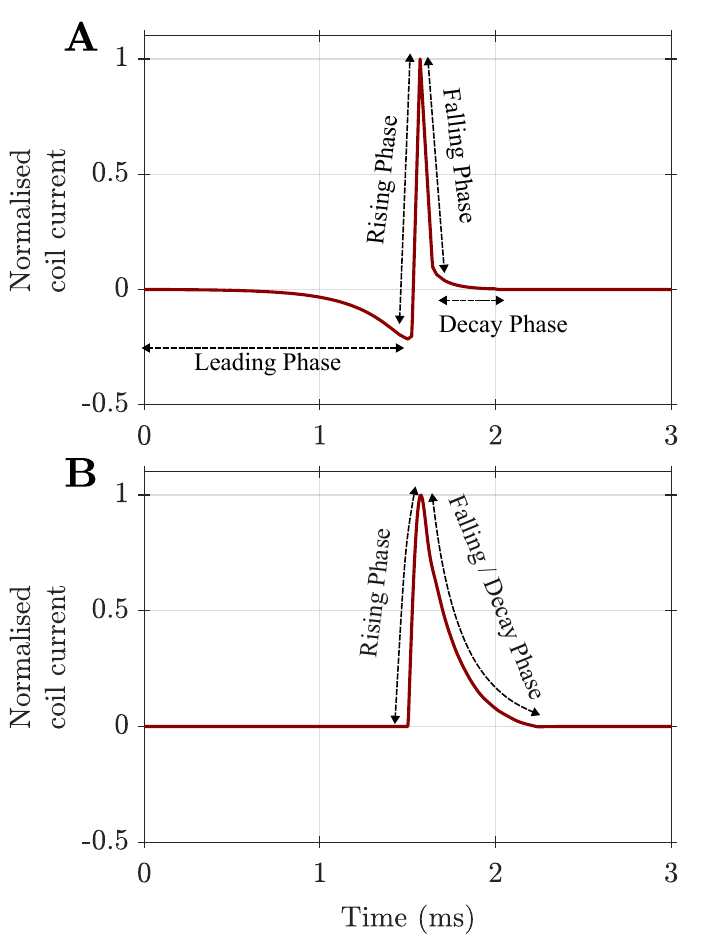}
    \caption{\textbf{An optimised and conventional current waveforms.} (\textbf{A}) Optimised current waveforms typically have four distinct phases, specifically a leading phase, a rising phase, a falling phase, and a decay phase. The shown waveform has voltage limits of $V_\textrm{max}=2000$ and $V_\textrm{min}=-250$. (\textbf{B}) The current waveform of conventional monophasic pulses has no leading phase and the falling and decay phases are not differentiated as in the optimised waveform. This waveform was computed by integrating the electric field waveform recorded from a MagPro X100 device (MagVenture A/S, Farum, Denmark). Both current waveforms are normalised to 1 at their peak amplitude.}
    \label{fig: representative current waveform}
\end{figure}

\begin{figure}
    \centering
    \includegraphics[width=\linewidth]{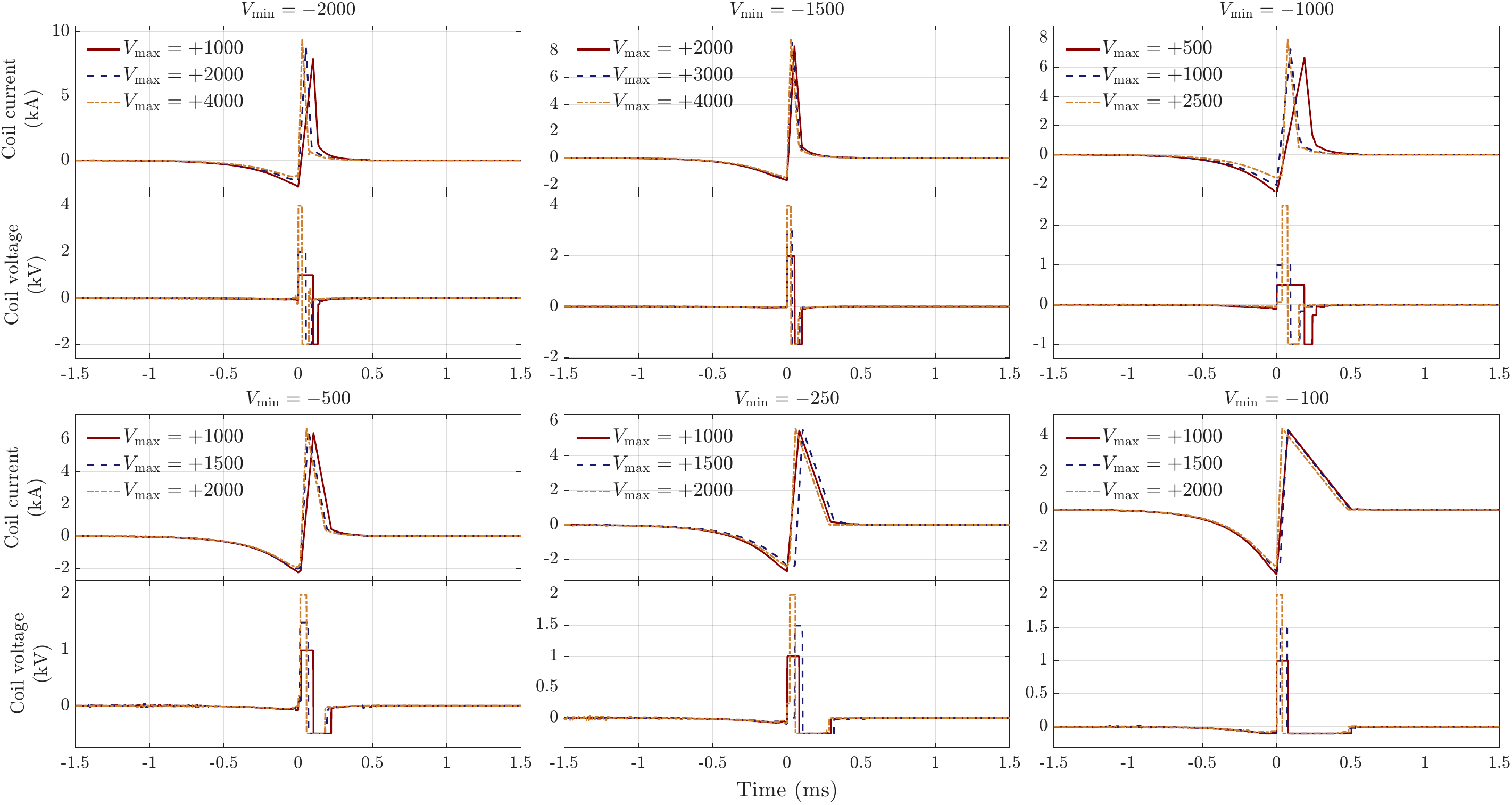}
    \caption{\textbf{Optimised current waveforms (top row) and corresponding voltage (electric field) waveforms (bottom row) for various asymmetry levels.} Each column shows the waveforms with a given negative voltage limit and several positive voltage limits. All coil voltage waveforms are nearly rectangular with different pulse amplitudes and widths.}
    \label{fig: Optimised waveforms}
\end{figure}

A representative optimised current waveform shown in Figure \ref{fig: representative current waveform}(\textbf{A}) demonstrates shared features found in the optimised waveforms (Figure \ref{fig: Optimised waveforms}). The current waveform of a conventional monophasic pulse has only a positive current direction, with both rising and falling phases (Figure \ref{fig: representative current waveform}(\textbf{B})). We divided an optimised current waveform sequentially into four distinct phases for later analysis. It begins with a leading phase, where the current gradually decreases from its starting point to a minimum negative value. This baseline shift is immediately followed by a rising phase, characterised by a rapid increase from this lowest point to the maximum positive amplitude. Subsequently, the waveform enters a falling phase, during which the current decreases approximately linearly from its peak to an intermediate value. The final component is the decay phase, where the current slowly diminishes from this intermediate value and eventually returns to zero. This sequence of phases---initial decrease, swift rise, linear fall, and gradual decay---collectively describes the complete evolution of the current waveform over time. The corresponding voltage and electric field waveforms have small negative amplitudes during the first and fourth phases, and unequal positive and negative phases during the second and third phases, respectively, corresponding to the ascending and descending slopes for the rising and falling phases in current waveforms.

\subsection{Analysis of waveform parameters}
The detailed waveform parameters are listed in Table S2 (see Supplementary Materials). The voltage limits reflect the upper and lower levels ${V}_\textrm{max}$ and ${V}_\textrm{min}$ used in the penalty term in the objective function (Table \ref{tab: optimisation problem}). $\hat{V}_\textrm{max}$ and $\hat{V}_\textrm{min}$ are respectively the maximum and minimum voltages of the optimised results. We defined the effective pulse duration (full width at half maximum), $T_\textrm{pulse}$, as the sum of the second (rising phase, $T_\textrm{rise}$) and third (falling phase, $T_\textrm{fall}$) current phases, where the corresponding coil voltage reached at least $50\,\%$ of the true maximum $\hat{V}_\textrm{max}$ or minimum $\hat{V}_\textrm{min}$ coil voltages. $\hat{I}_\textrm{max}$ and $\hat{I}_\textrm{min}$ are respectively the maximum and minimum amplitudes of the optimised current waveforms. $\tau_\textrm{init}$ is the time constant of a fitted exponential function, $I(t) = \hat{I}_\textrm{min} \cdot \exp\big((t-T_\textrm{init}) / \tau_\textrm{init}\big)$, for the leading phase of the current waveforms. $T_\textrm{init}$ is the leading phase duration that is defined as the time for the current waveform decreasing from $I(0)=0$ to the point where $I(T_\textrm{init}) = \hat{I}_\textrm{min}$. Moreover, we introduce the current ratio as $r_\textrm{I} = \abs{\frac{\hat{I}_\textrm{max}}{\hat{I}_\textrm{min}}}$ to evaluate the optimised current waveforms and used a power function $f(X) = a\cdot X^b$ or a logarithm function $f(X) = a\cdot \log(X) + b$ to investigate the correlations among the pulse duration (in units of $\textrm{ms}$), the current and voltage waveforms, and the energy loss $\hat{\mathcal{W}}$. Finally, we evaluated coefficients of determination ($R^\textrm{2}$) for each regression.

\begin{figure}
    \centering
    \includegraphics[width = 0.8\linewidth]{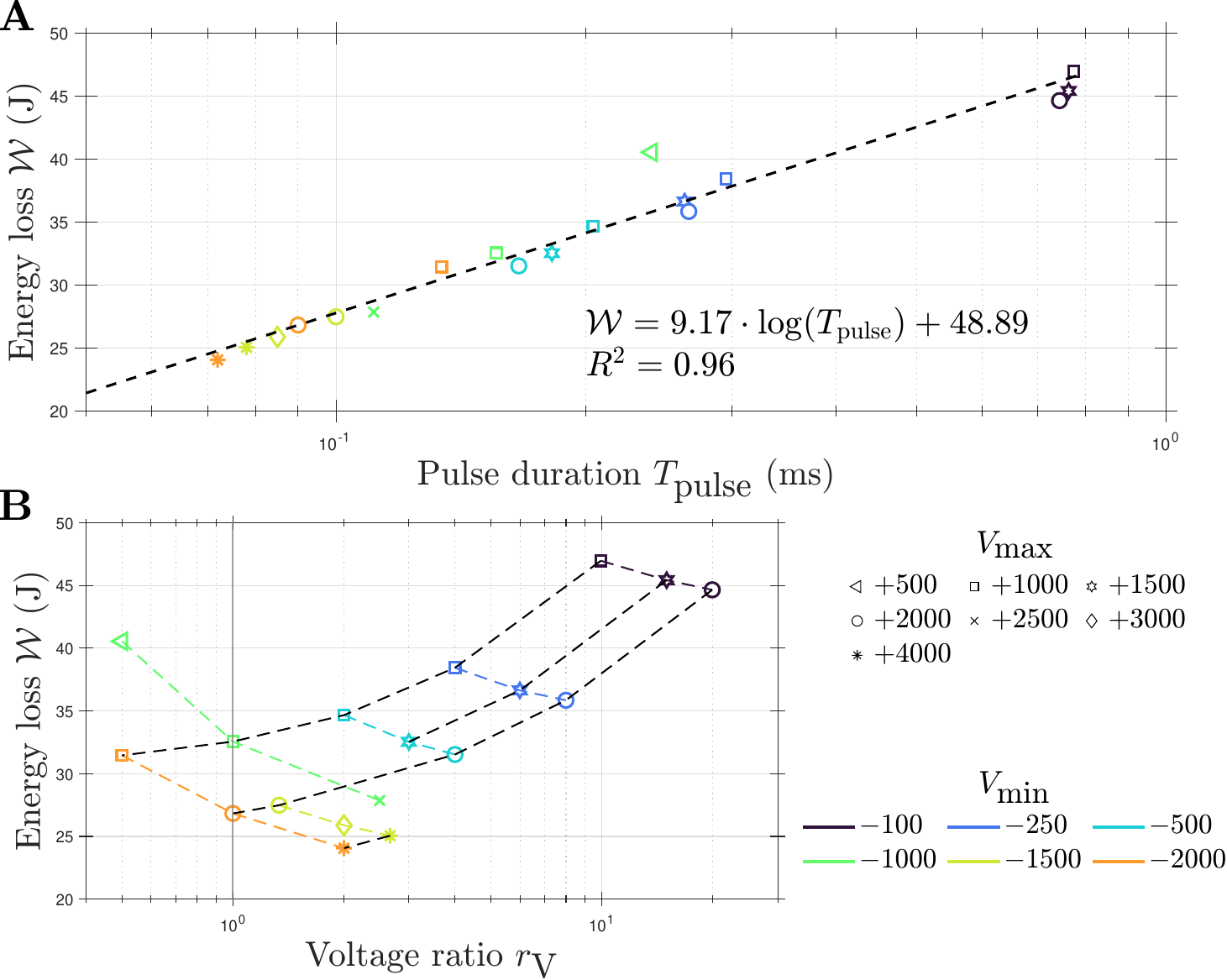}
    \caption{\textbf{Energy loss versus pulse duration (A) and voltage ratio (B) of the optimised pulses.} Marker shape and colour correspond to specific positive and negative voltage limits, respectively. $\mathcal{W}$ represents simulated energy loss in joules. The high coefficients of determination (${R}^\textrm{2} = 0.96$) suggest that the regression curve accurately describes the trends.}
    \label{fig: energy cost}
\end{figure}

\begin{figure}
    \centering
    \includegraphics[width = \linewidth]{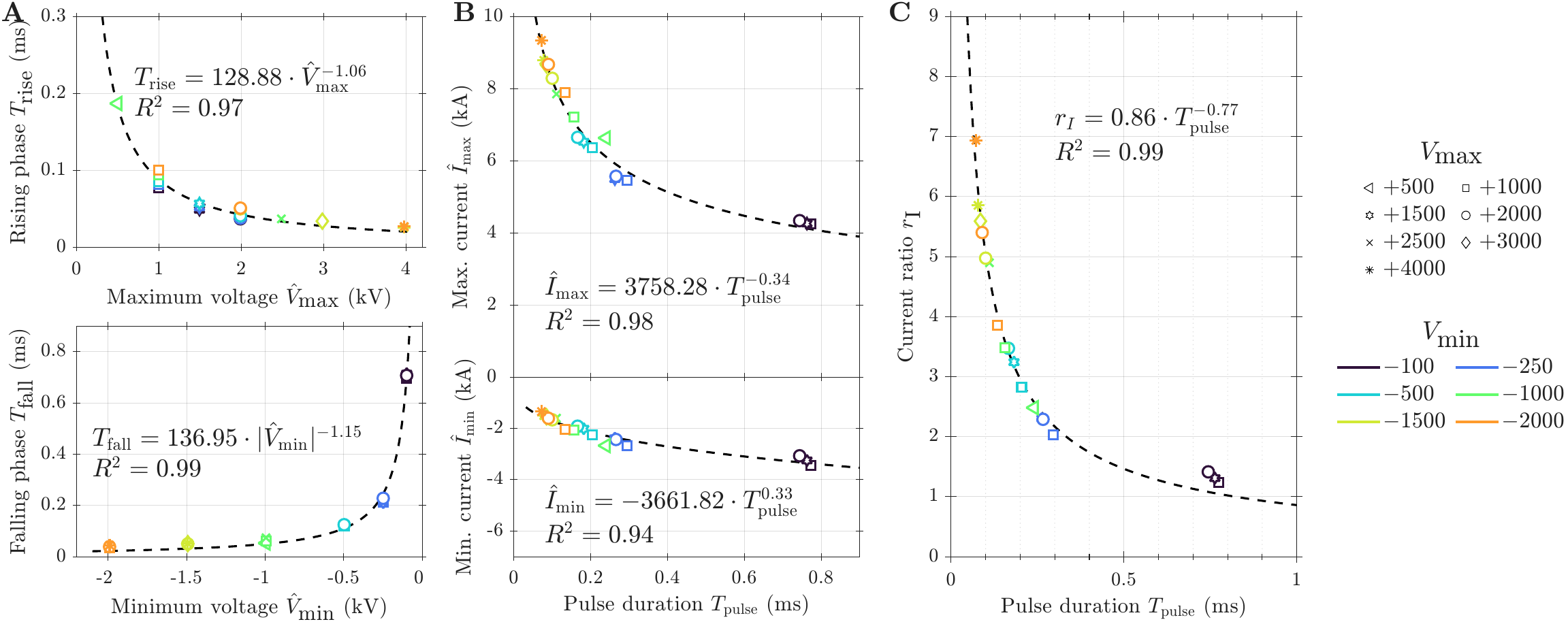}
    \caption{\textbf{Relationships between the parameters of the optimised waveforms.} (\textbf{A}) Duration of the rising (top) and falling (bottom) phases of the current waveform, respectively, versus the positive and negative voltage limits. (\textbf{B}) Maximum (top) and minimum (bottom) current amplitudes versus the pulse duration. (\textbf{C}) Current asymmetry ratio $r_\textrm{I}$ versus the pulse duration. The generation of symmetric current pulses (though still asymmetric in voltage and electric field) can have technological advantages. The high coefficients of determination ($R^\textrm{2} \geq 0.94$) suggest that the regression curves accurately describe the trends.}
    \label{fig: data analysis}
\end{figure}

{
The optimised current waveforms had similar decay constants of 280\,$\pm$\,15\,{\textmu}s (mean\,$\pm$\,SD) for the leading phase (${R}^\textrm{2} \geq 0.98$ for all fitted exponential functions). The corresponding voltage waveforms (Figures \ref{fig: Optimised waveforms}) exhibited near-rectangular shapes with varying pulse widths and asymmetry levels. The analysis shown in Figure \ref{fig: energy cost}(\textbf{A}) revealed that the energy loss increases logarithmically with pulse duration. Figure \ref{fig: energy cost}(\textbf{B}) shows that the energy losses for pulses with a fixed positive phase increase as the voltage ratio increases, while those with a fixed negative phase decrease as the voltage ratio increases. These patterns suggest that higher voltage amplitudes lead to shorter pulses in both positive and negative voltage phases, as expected \cite{peterchev2013pulse}. Notably, Figure \ref{fig: energy cost}(\textbf{B}) also demonstrates that some optimised asymmetric pulses could have superior power efficiency compared to symmetric ones when higher voltages and shorter pulses were permitted. For instance, the waveforms V:+4000/--2000 and V:+4000/--1500 achieved lower energy loss than V:+2000/--2000.
}

Further investigation of the relationship between pulse width and voltage amplitude, shown in Figure \ref{fig: data analysis}(\textbf{A}), demonstrated that the durations of both the rising $T_\textrm{rise}$ and falling $T_\textrm{fall}$ phases decrease with increasing pulse magnitude, with the fitting curves describing these trends well (${R}^\textrm{2} \geq 0.97$). Moreover, the optimised results revealed several monotonic relationships between current waveforms and pulse duration. Figure \ref{fig: data analysis}({\bf B}) shows that as the pulse duration increases, the peak current $\hat{I}_\textrm{max}$ at the threshold decreases, whereas the magnitude of the leading current phase $\abs{\hat{I}_\textrm{min}}$ increases ($R^2 \geq 0.94$). Thus, longer rising phases and pulse durations require deeper leading current phases for compensation. Figure \ref{fig: data analysis}({\bf C}) illustrates the inverse relationship between pulse duration and current ratio ($R^2=0.99$); larger current ratios correspond to shorter pulse durations. This relationship is also reflected in the upward shift of the rising phases as pulse duration decreases (visible in Figure \ref{fig: Optimised waveforms}). These findings suggest an optimisation strategy: when the current amplitude is high, shorter phase durations minimise energy cost. This relationship creates a cascading effect where larger current ratios lead to shorter pulse durations, which ultimately reduces energy loss.

\begin{figure}
    \centering
    \includegraphics[width=\linewidth]{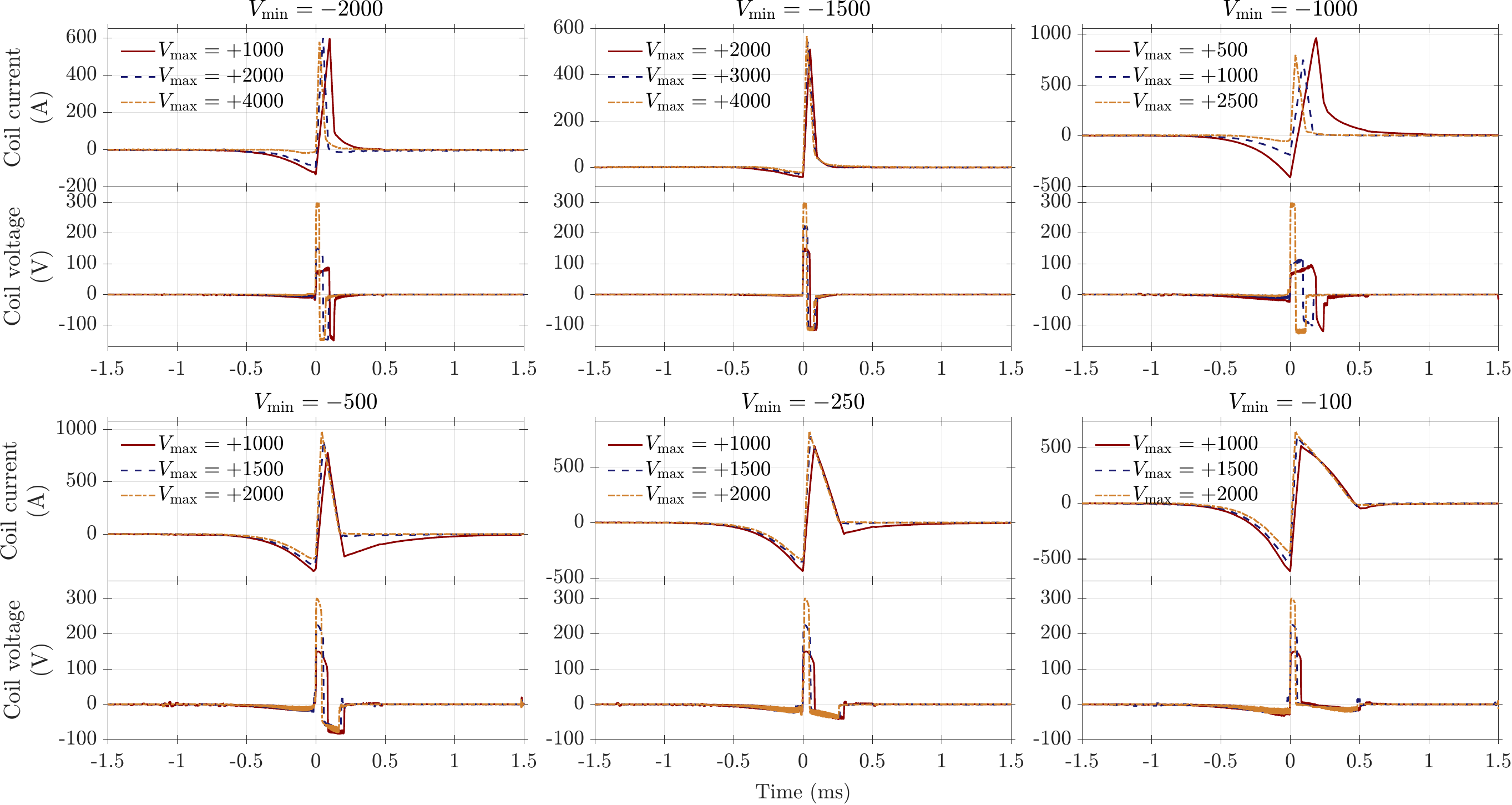}
    \caption{\textbf{Experimental coil current and voltage traces of the optimised waveforms generated by MPS-TMS.} For visualisation purposes, both current and voltage waveforms in each $V_\textrm{min}$ group were filtered to remove measurement noise and proportionally scaled according to the amplitude ratio between them.}
    \label{fig: recorded waveforms}
\end{figure}

\subsection{Comparison of energy loss}
\begin{figure}
    \centering
    \includegraphics[width=\linewidth]{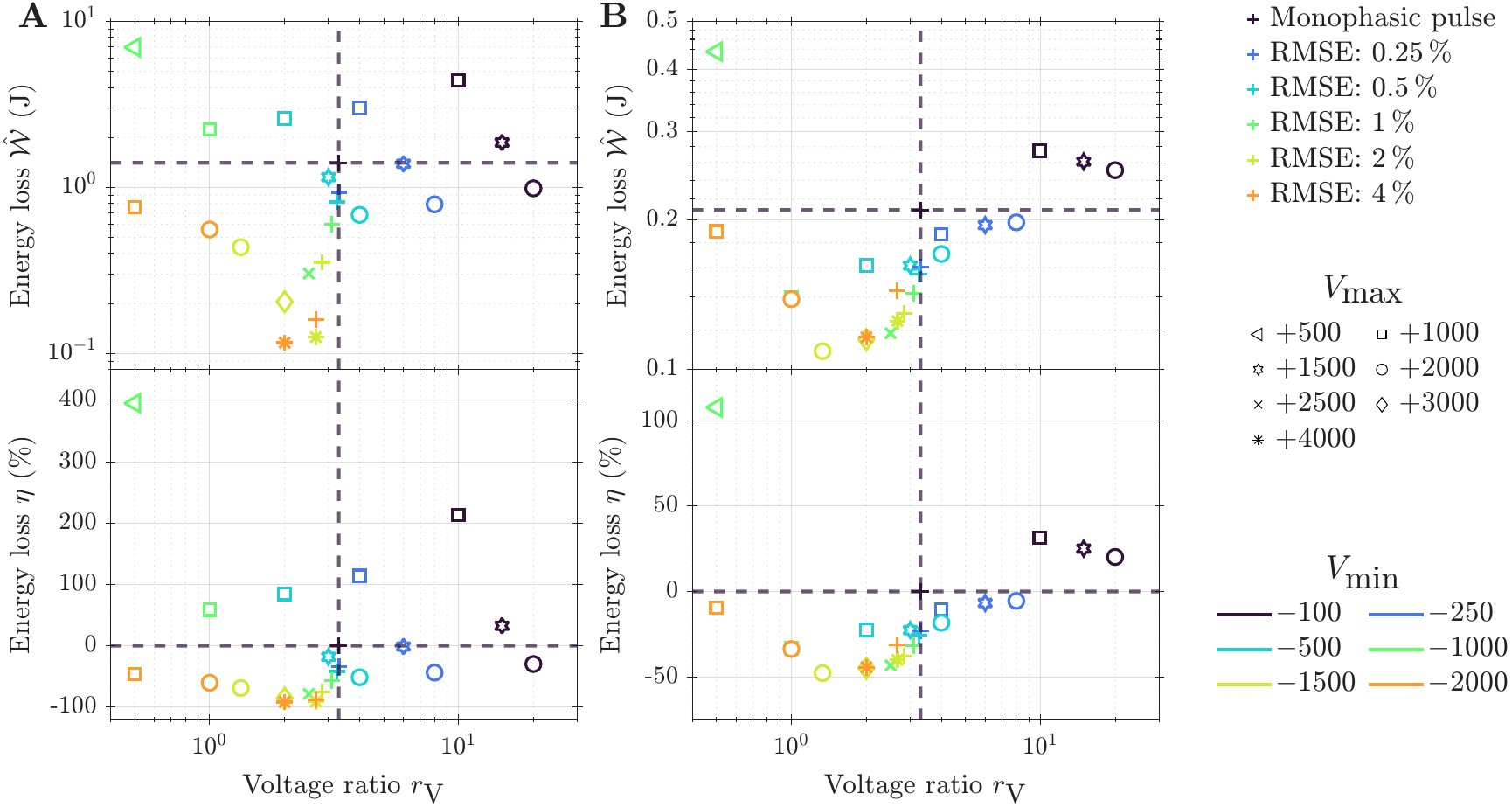}
    \caption{\textbf{Ohmic heating of the coil ($\hat{\mathcal{W}}$) for optimised pulses with various asymmetry ratios compared to the conventional monophasic and monophasic-equivalent pulses.} (\textbf{A}) Pulses with the same maximum voltage for peak-matched comparison. (\textbf{B}) Pulses with the scaled amplitudes for threshold-matched comparison. The upper panels show the absolute energy losses and the lower panels present the relative changes compared to the conventional monophasic pulse. The vertical and the horizontal dashed red lines respectively represent the asymmetry ratio and the corresponding energy loss/percentage of the conventional monophasic pulse. $\hat{\mathcal{W}}$ represents the experimental energy loss.}
    \label{fig: energy loss}
\end{figure}

The current waveforms generated with the MPS-TMS device (Figure \ref{fig: recorded waveforms}, upper rows) were generally faithful replications of the computational optimisation, and the corresponding voltage waveforms were almost near-rectangular with various voltage ratios satisfying the constraints of the optimisation (lower rows). In the peak-matched comparison, the majority of optimised asymmetric pulses ($12$ out of $18$) demonstrate significantly lower losses up to $92\,\%$ less than conventional monophasic pulses, while some of these pulses with higher losses always have $V_\mathrm{max} \leq 1500$. In addition, the losses of optimised pulses can achieve up to $88\,\%$ less than optimised monophasic-equivalent pulses (Figure \ref{fig: energy loss}(\textrm{A}), absolute losses in the upper panel and relative changes compared to conventional monophasic pulse in the lower panel).

After scaling according to their respective simulated neural thresholds, $14$ out of $18$ threshold-matched optimised asymmetric pulses generated less losses than the conventional monophasic pulse (Figure \ref{fig: energy loss}(\textbf{B})). For example, the optimised asymmetric pulse \textrm{V:+2000/--1500} resulted in $55\,\%$ less heating compared to the conventional monophasic pulse and $16\,\% - 31\,\%$ less heating than monophasic-equivalent pulses. Moreover, the optimised asymmetric pulses in both peak- and threshold-matched cases can achieve lower energy losses with higher asymmetry ratios than conventional monophasic and monophasic-equivalent pulses.

{
The experimental validation revealed some variability in energy reduction compared to the computational predictions (Figure \ref{fig: energy loss}). This deviation is primarily attributed to waveform distortion during the experimental synthesis process. We calculated the correlation coefficients $R$ between the simulated and experimental current waveforms; most of them have $R>0.96$, while \textrm{V:+500/--1000} has the lowest value of $R = 0.91$. Although the MPS-TMS device successfully generated most optimised waveforms with high fidelity, slight waveform distortions directly impact energy performance, as demonstrated by pulse \textrm{V:+500/--1000} (the worst case, green triangle in Figure \ref{fig: energy loss}).
}

\subsection{Resting motor threshold, energy loss, and MEP latency}
\begin{figure}
    \centering
    \includegraphics[width=0.7\linewidth]{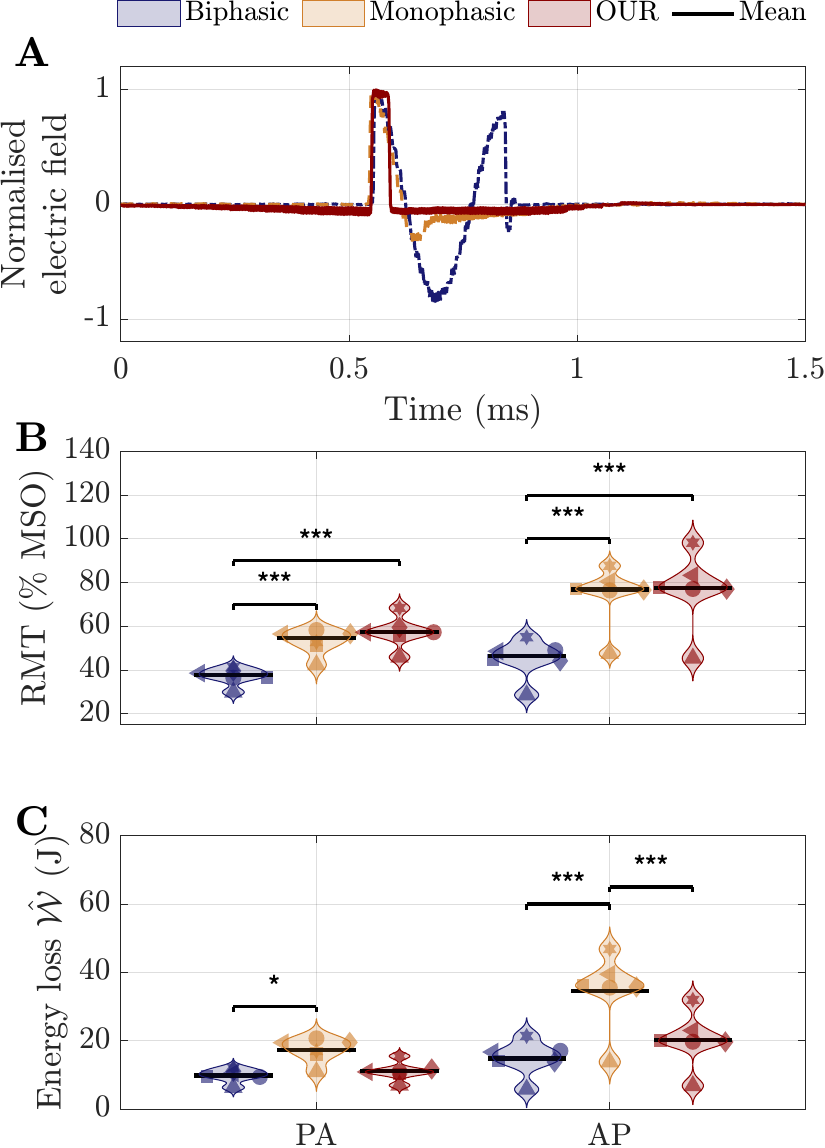}
    \caption{\textbf{Recorded waveforms used in human experiments along with their associated RMTs and energy losses.} (\textbf{A}) The normalised recorded electric field waveforms of optimised and conventional pulses. (\textbf{B}) RMTs statistics of all subjects assessed by these pulses in both posterior--anterior (PA) and anterior--posterior (AP) directions. (\textbf{C}) The corresponding energy losses of the pulses at their RMTs. Different markers denote different subjects. Markers with blue, yellow, and red respectively, are for biphasic, monophasic and OUR pulses. OUR pulse refers to the optimised waveform \textrm{V:+2000/--100} with $r_\textrm{V} = 20$. For visualisation purposes, the electric field waveforms were low-pass filtered to remove measurement noise. * $p < 0.05$, *** $p < 0.001$.}
    \label{fig: human waveforms}
\end{figure}

{No EMG afterdischarges or spread of muscle excitation were observed and no unexpected or serious adverse events occurred during the human subjects study.} Figure \ref{fig: human waveforms}(\textbf{A}) illustrates the pulse waveforms used in the human subjects study. Figures \ref{fig: human waveforms}(\textbf{B}) and (\textbf{C}) respectively present RMTs and energy losses of the different pulses. The mixed-effects model revealed significant main effects of \textit{pulse shape} ($F(2, 25) = 70.9$, $p < 0.001$), \textit{pulse direction} ($F(1, 25) = 69.8$, $p < 0.001$), and their interaction ($F(2, 25) = 4.64$, $p = 0.019$) in RMTs. Post-hoc analyses demonstrated that in both directions, OUR and monophasic pulses required similar pulse amplitude (all $p > 0.78$), which was always higher than that of biphasic pulses (all $p < 0.001$). Moreover, the RMT differences between AP and PA directions of monophasic ($21.3\,\%\,\textrm{MSO}$, $p < 0.001$) and OUR ($19.1\,\%\,\textrm{MSO}$, $p < 0.001$) pulses were significant. In contrast, biphasic pulses demonstrated no significant RMT difference between PA and AP directions ($7.90\,\%\,\textrm{MSO}$, $p = 0.207$).

The energy loss at RMT for the different pulses is presented in Figure \ref{fig: human waveforms}(\textbf{C}). There were significant effects of \textit{pulse shape} ($F(2, 25) = 39.1$, $p < 0.001$), \textit{pulse direction} ($F(1, 25) = 63.9$, $p < 0.001$), and their interaction ($F(2, 25) = 7.66$, $p = 0.002$). In the AP direction, monophasic pulses had the highest energy loss ($34.6\,\textrm{J}$), followed by OUR pulses ($20.2\,\textrm{J}$), and then biphasic pulses ($14.8\,\textrm{J}$). The differences between monophasic and biphasic ($p < 0.001$) and monophasic and OUR ($p < 0.001$) were significant, while OUR did not significantly differ from biphasic ($p = 0.202$). This hierarchical pattern notably differs from the RMT results, where OUR and monophasic pulses showed similar thresholds in AP direction. In PA direction, the pulse energy losses were consistently lower than in the AP direction. There was no significant differences between biphasic ($9.84\,\textrm{J}$) and OUR ($11.1\,\textrm{J}$; $p = 0.994$). In contrast, the energy loss of biphasic pulses was significantly lower than that of monophasic pulses ($17.3\,\textrm{J}$; $p = 0.033$). The difference between monophasic and OUR did not reach statistical significance ($6.22\,\textrm{J}$; $p = 0.103$). The energy-loss data contrast with the RMT findings, where OUR and monophasic pulses had higher thresholds than biphasic pulses in PA direction.

\begin{figure}[!htbp]
    \centering
    \includegraphics[width=0.6\linewidth]{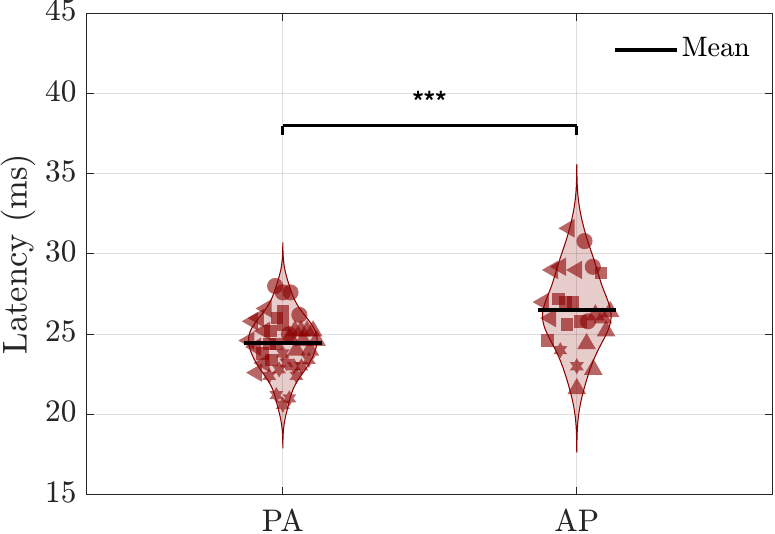}
    \caption{\textbf{MEP latencies statistics of OUR pulse in PA and AP directions.} OUR pulse refers to the optimised waveform \textrm{V:+2000/--100} with $r_\textrm{V} = 20$. Different markers denote different subjects. *** $p < 0.001$.}
    \label{fig: OUR MEP latency}
\end{figure}

Figure \ref{fig: OUR MEP latency} presents MEP latencies of OUR pulses for PA and AP directions. The mixed-effects model revealed significant main effect of \textit{pulse direction} ($F(1, 56.6) = 19.0$, $p < 0.001$). Post-hoc analysis demonstrated that OUR pulses produced a significant latency difference of $1.79\,\unit{\ms}$ ($p < 0.001$) depending on pulse direction: AP direction yielded a higher latency ($26.4\,\unit{\ms}$, $n = 25$) than PA direction ($24.7\,\unit{\ms}$, $n = 41$).

{
The cross-validation analysis demonstrated consistency of our statistical findings across all measurements (see Supplementary Material, Section 5). Main effects for pulse shape and direction remained highly significant across all cross-validation folds for both the RMT and energy loss analyses (all $p < 0.001$). MEP latency directional selectivity similarly showed consistent significance across all five available folds (all $p < 0.05$). These results confirm that our findings are not dependent on any individual participant or driven by outliers.
}

\section{Discussion}
This study proposed a new optimisation framework that considers the activation of a computational neuron model as a constraint and numerically optimises the TMS pulse shape, particularly of selective asymmetric pulses, to minimise energy loss and coil heating represented by the integral of the squared coil current. The optimisation process was not limited to specific devices or predetermined waveform parametrisation. Instead, we developed a global search of the pulse shape space for various asymmetry levels and pulse durations. We constrained the optimisation by limiting the maximum and minimum coil voltages and ensuring that the neuron model can generate an action potential.

\subsection{Waveform energy loss}
The optimisation generated asymmetric near-rectangular electric fields with near-triangular current shapes with specific features. Conventional biphasic and monophasic TMS devices, on the other hand, induce under-damped or damped cosinusoidal electric field waveforms due to circuit limitations. The optimal current waveforms we identified consisted of an initially slowly falling phase followed by rapidly rising and falling phases, and trailed by a slow decay to zero. These features share similarities with the optimisation results for purely symmetric pulses \cite{goetz2013_biphasic_opt}; specifically, the leading phase shifted the current baseline downwards, which also evolved in the optimisation of monophasic-equivalent pulses \cite{wang2022optimized}.

The current in the leading phase shifted the baseline so slowly---i.e., had a small time derivative---that this phase hardly showed up in the electric field. The leading phase followed a near-exponential course with a relatively consistent time constant of around {(280\,$\pm$\,15)\,{\textmu}s} (mean $\pm$ SD) across all optimised current waveforms, which is not far above typical axonal time constants \cite{peterchev2013pulse, dostilio2016effect, barker1991introduction}. The depth of the current leading phase can be substantial in magnitude and even reach the same value as the positive current peak, which can have advantages in implementation. Since this feature emerges time and again in pulse-shape optimisation, it appears that the current leading phase may be a universal feature of coil-heating-optimised pulses of any shape and could be added to pulse-shape designs provisionally without running computationally expensive numerical optimisation \cite{goetz2013_biphasic_opt, goetz2012optimization,wang2022optimized,WO2013131639A1}.

Moreover, our optimisation framework forced the optimised pulse to stay at the desired asymmetry ratio for activation selectivity and ensured low energy loss. In contrast, relaxing this constraint has previously reduced the energy loss of optimised monophasic-equivalent pulses but also their corresponding asymmetry ratios by introducing a greater depth in the current leading phase \cite{wang2022optimized}. For example, the pulse with an RMSE deviation of $4\,\%$  from conventional monophasic exhibited an asymmetry ratio of $r_\textrm{V} = 2.7$, which is lower than that of conventional monophasic pulses ($r_\textrm{V} = 3.3$).

Further, our findings confirmed that regardless of whether an electric field pulse is asymmetric or symmetric, shorter pulses reduce coil heating and losses in the device but require higher voltages to reach the stimulation threshold. This observation is in line with previous results for sinusoidal pulses \cite{goetz2013_biphasic_opt,cadwell1991optimizing,barker1991_Pulse_Duration_Amplitude} and asymmetric rectangular pulses \cite{goetz2012optimization,peterchev2011repetitive}.

{
Our analysis focused primarily on the coil resistance (generally the total resistance in the current path) as the dominant energy loss factor and pivotal for coil heating. Other waveform-dependent losses include switching losses in the semiconductor device(s), which depend on the voltage as well \cite{peterchev2011repetitive,Zeng2022,li2022modular}. The switching losses are generally much smaller than the conduction (resistance-based) losses and depend strongly on the specific pulse generator implementation. It should be noted that the device switching losses do not contribute to the coil heating. Further, coils with a ferromagnetic core can exhibit core losses, although they are typically also smaller than the conduction losses \cite{lorenzen1992,epstein2002iron}. If any of these losses are significant for a specific hardware design, they can be incorporated in our optimisation framework by adding extra terms to the cost function.
}

\subsection{Waveform activation selectivity}
The optimised asymmetric pulses are designed to preserve or increase the selective stimulation properties characteristic of conventional monophasic pulses and still reduce their large energy loss. Stimulation selectivity is typically demonstrated by direction-dependent effects, where different current directions or coil orientations produce distinct physiological responses. Previous studies demonstrated that asymmetric pulses, compared to symmetric ones, generally exhibit significantly greater differences in motor thresholds and MEP latencies when the stimulation direction (AP vs.\ PA) is reversed \cite{sommer2006_waveform_motor_cortex,goetz2016_Enhancement_RecMono}. To validate this selectivity for optimised pulses, we tested OUR pulses (\textrm{V:+2000/--100}) in healthy human subjects. OUR pulses demonstrated significant directional dependence in both RMT and MEP latency between AP and PA directions. With respect to RMTs, OUR pulses had similar performance with monophasic pulses. Notably, the AP vs.\ PA MEP latency mean difference of OUR pulses of $1.79\,\unit{\ms}$ is within the range of previous studies with conventional monophasic pulses ($0.9-1.9$\,ms) \cite{sommer2006_waveform_motor_cortex, ni2011_mep_latency,laakso2025_MEP_latency, davila-perez_effects_2018, jo2018_MEP_latency}. Further, OUR pulse had a brief main phase of $40\,${\textmu}s and was highly asymmetric, both of which could be advantages for selective neural population stimulation \cite{dostilio2016effect, sommer2018_cTMS_biphasic, di2001_biphasic_volleys}.

{
\subsection{Clinical and practical implications}
Conventional TMS devices typically manage excessive energy dissipation and coil heating by a combination of cooling systems and restrictions on the pulse amplitude, repetition frequency, and/or treatment duration \cite{weyh2005_TMS_heating_equation, zacharias2019method}. These limitations preclude, for example, monophasic pulses from being deployed in high-frequency protocols with conventional rTMS devices. In contrast, our approach reduces energy dissipation and heating through more efficient pulse shapes, which obviates the need for complex cooling systems and expensive high-capacity power supplies \cite{nielsen1995_oil_cooled_coil, weyh2005_TMS_heating_equation}. The direction-dependent MEP latency suggests selective stimulation of specific neural circuits, which could potentially enhance neuromodulation and hence therapeutic applications \cite{sommer2002_mono_effective, arai2007_mono_effective, goetz2016_Enhancement_RecMono, wendt2023_iTBS_monoVSbi, Sommer2024}. The reported small human subjects study was designed and powered to confirm the large effect of pulse shape and direction on energy dissipation and MEP latency in a single-pulse paradigm; subsequent studies could explore the impact of these pulse characteristics on neuromodulation protocols.  
}

\section{Conclusion}
This study presents a novel optimisation framework that successfully identifies energy-efficient asymmetric TMS pulse shapes with potentially high neural activation selectivity. Unlike previous work that focused on symmetric pulses or was constrained to match conventional monophasic pulses, our minimally-constrained optimisation approach revealed that near-rectangular electric field pulses with specific current features can reduce coil heating by up to 92\,\% compared to conventional monophasic pulses. The optimised asymmetric pulses demonstrated strong directional dependence of motor threshold and MEP latency, which suggests selective stimulation capabilities. These highly efficient pulses can enable selective rapid-rate TMS protocols with reduced power consumption and coil heating and can expand the applicability of asymmetric pulses in neuromodulation applications.

\section*{Author Contributions}
SMG conceived and supervised the computational study and developed the initial method as well as the original optimisation code for asymmetric pulses. AVP supervised the experimental study. SMG and AVP secured funding for the study. KM refined the computational method as well as the code and performed numerical optimisation, simulation result collection, data analysis on simulated and human data, and visualisation. AV, ZBS, DLKM, and YL conducted the human experiments. JZ implemented modular pulse synthesiser hardware as well as software and recorded pulse waveforms. BW performed data analysis on partial simulated data and visualisation. YL and DLKM implemented the experimental measurement procedure. JYC provided regulatory and subject recruitment support. MEC provided technical support for the experiments. NBP was the study physician. KM and SMG wrote the manuscript, BW and AVP edited the text, and all authors reviewed, commented on, and approved the final version of the manuscript.

\section*{Acknowledgments}
This work was supported by Trinity College's Isaac Newton Trust and the US National Institute of Mental Health (Grant No.\ 1RF1MH124943). The content is solely the responsibility of the authors and does not necessarily represent the official views of the funders. Preliminary results were presented at the \textit{6\textsuperscript{th} International Brain Stimulation Conference}, Kobe, Japan, Feb. 23–26, 2025 \cite{peterchev2025experimental, vlasov2025first}.

\section*{Conflicts of interests}
S.~M.~Goetz is inventor on patents and patent applications on MPS-TMS, related power electronics circuits, and other TMS technology. Related to TMS technology, he has previously received royalties from Rogue Research as well as research funding from Magstim. A.~V.~Peterchev is an inventor on patents and patent applications on transcranial magnetic stimulation technology and has received patent royalties and consulting fees from Rogue Research; equity options, scientific advisory board membership, and consulting fees from Ampa Health; equity options and consulting fees from Magnetic Tides; consulting fees from Soterix Medical; equipment loans from MagVenture; and research funding from Motif. Other authors declare no relevant conflict of interest.

\printbibliography
\end{document}